\documentclass[a4paper,11pt, onesided]{article}

\usepackage[ansinew]{inputenc}

\usepackage{amsfonts}
\usepackage{amsmath}
\usepackage{amsthm}

\usepackage{mathrsfs}
\usepackage{amssymb}

\newtheorem{theorem}{Theorem}
\newtheorem{lemma}[theorem]{Lemma}

\begin{document} 

\title{Conserved Matter Superenergy Currents for Hypersurface
  Orthogonal Killing Vectors}
\author{Ingemar Eriksson \\
       Matematiska institutionen, Link\"opings universitet, \\
       SE-581 83 Link\"oping, Sweden \\
       ineri@mai.liu.se}
\date{ November 2, 2005}

\maketitle
\abstract{ We show that for hypersurface orthogonal Killing vectors, the
corresponding Chevreton superenergy currents will be conserved and
proportional to the Killing vectors. This holds for four-dimensional
Einstein-Maxwell spacetimes with an electromagnetic field that is
source-free and inherits the symmetry of the spacetime. A similar
result also holds for the trace of the Chevreton tensor.
The corresponding Bel currents have previously been proven
to be conserved and our result can be seen as giving further support to
the concept of conserved mixed superenergy currents. The analogous case
for a scalar field has also previously been proven to give conserved currents and
we show, for completeness, that these currents also are proportional
to the Killing vectors. }

\section{Introduction}

The Bel-Robinson tensor was introduced in 1958 as an attempt to
describe gravitational energy \cite{Bel1958}, \cite{Bel1959}. It is given by
\begin{align} \label{BRTensor}
  T_{abcd} =& C_{aecf}C_b{}^e{}_d{}^f + C_{aedf}C_b{}^e{}_c{}^f
           - \frac{1}{2}g_{ab}C_{efcg}C^{ef}{}_d{}^g\nonumber\\
           &- \frac{1}{2}g_{cd}C_{aefg}C_b{}^{efg}
           + \frac{1}{8}g_{ab}g_{cd}C_{efgh}C^{efgh},
\end{align}
where $C_{abcd}$ is the Weyl tensor. It is completely symmetric in
four and five dimensions and it is divergence-free in vacuum \cite{Senovilla2000}. 
The Bel tensor was introduced shortly afterwards as an 
extension to non-vacuum cases, 
\begin{align}
  B_{abcd} =& R_{aecf}R_b{}^e{}_d{}^f + R_{aedf}R_b{}^e{}_c{}^f
           - \frac{1}{2}g_{ab}R_{efcg}R^{ef}{}_d{}^g\nonumber\\
           &- \frac{1}{2}g_{cd}R_{aefg}R_b{}^{efg}
           + \frac{1}{8}g_{ab}g_{cd}R_{efgh}R^{efgh},
\end{align}
where $R_{abcd}$ is the Riemann tensor. It has the following
symmetries, $B_{abcd} = B_{(ab)(cd)} = B_{cdab}$. Both these tensors
are divergence-free for Einstein spaces, $R_{ab} = \Lambda g_{ab}$,
but neither of them is divergence-free in general \cite{Senovilla2000}.

In 1964 Chevreton \cite{Chevreton1964} introduced an analogous tensor for the electromagnetic
field, the Chevreton tensor,
\begin{align}
     H_{abcd} = 
  & -\frac{1}{2}( \nabla_a F_{ce} \nabla_b F_d{}^e + 
                             \nabla_b F_{ce} \nabla_a F_d{}^e +
                             \nabla_c F_{ae} \nabla_d F_b{}^e +
                             \nabla_d F_{ae} \nabla_c F_b{}^e ) \nonumber\\
             & +\frac{1}{2}( g_{ab}\nabla_f F_{ce} \nabla^f F_d{}^e +
                             g_{cd}\nabla_f F_{ae} \nabla^f F_b{}^e)
              +\frac{1}{4}( g_{ab}\nabla_c F_{ef} \nabla_d F^{ef} +
                             g_{cd}\nabla_a F_{ef} \nabla_b F^{ef})\nonumber\\
             & -\frac{1}{4} g_{ab}g_{cd}\nabla_e F_{fg} \nabla^e F^{fg},
\end{align}
where $F_{ab}$ is a Maxwell field. In source-free regions, this tensor is
completely symmetric in four dimensions \cite{Bergqvist2003} and it is
divergence-free in flat spacetimes.

These three tensors are referred to as superenergy tensors, since they
have properties similar to the ordinary energy-momentum tensors, but
they do not have units of energy density. It seems rather that the correct
interpretation is units of energy density per unit surface.
These tensors can now be seen as special cases of a general
superenergy
tensor construction in Lorentzian manifolds of arbitrary dimension
\cite{Senovilla2000}.
All superenergy tensors, $T_{a_1\ldots
a_k}$, satisfy the Dominant Property, for $v_1,\ldots, v_k$
future-pointing causal vectors, $T_{a_1\ldots a_k}v_1^{a_1}\cdots
v_k^{a_k} \geq 0$ \cite{Bergqvist1999}, \cite{Senovilla2000}.

The Bel and Bel-Robinson tensors are not divergence-free in general
non-vacuum and the Chevreton tensor and the superenergy tensor of the scalar
field are not divergence-free in curved spacetimes. However, Senovilla
\cite{Senovilla2000}, \cite{Senovilla1999b}, has shown that for the Einstein-Klein-Gordon
theory, when a Killing vector, $\xi_a$, is present one can construct a 
conserved mixed current,
\begin{align}
 \nabla^a \left( (B_{abcd}+S_{abcd})\xi^b\xi^c\xi^d \right ) = 0, 
\end{align}
where $S_{abcd}$ is the superenergy tensor of the scalar field.
For the Einstein-Maxwell theory a similar result has been obtained for
propagation of discontinuities, but not in general
\cite{Senovilla1999a}, \cite{Senovilla2000}.

However, Lazkoz, Senovilla, and Vera \cite{Lazkoz2003} have shown that
for certain types of Killing vectors, the Bel currents will in fact be
independently conserved, which will prevent the interchange of
superenergy between the gravitational field and the matter fields in
such cases. This happens for hypersuface orthogonal Killing vectors
and for two commuting Killing vectors that act orthogonally transitive
on non-null surfaces. In the first case the corresponding Bel current
is proportional to the Killing vector, $\xi^a$, and in the second case the
currents are tangent to the Killing vectors $\xi_1^a$, $\xi_2^a$,
\begin{align} \label{Lazkoz:eq}
  B_{abcd}\xi^b\xi^c\xi^d &= \gamma\xi_a, & 
  B_{a(bcd)}\xi_i^b\xi_j^c\xi_k^d &= \alpha_{ijk}\xi_1{}_a 
                                  +\beta_{ijk}\xi_2{}_a.
\end{align}

In this paper we show that in the case of a hypersurface orthogonal
Killing vector in Einstein-Maxwell theory, if the electromagnetic
field is source-free and inherits the symmetry of the spacetime, then 
the Chevreton tensor will be independently conserved and proportional
to the Killing vector, giving further support to the concept of
conserved mixed superenergy currents. We already know that in the case
of Einstein-Klein-Gordon theory, the superenergy current of the scalar
field is conserved, but for completeness we also show that this
current is proportional to the Killing vector.

We will assume that our spacetime is four-dimensional and equipped
with a metric of signature $-2$. We define the Riemann
tensor as
\begin{align}
  (\nabla_a\nabla_b-\nabla_b\nabla_a)v_c = -R_{abcd}v^d.
\end{align}
The Einstein equations will be given as
\begin{align}
  R_{ab} - \frac{1}{2}R g_{ab} + \Lambda g_{ab} = -T_{ab}.
\end{align}
Note that we will keep the cosmological constant, $\Lambda$, through-out
the calculations. If $\xi_a$ is a Killing vector, then $\nabla_a \xi_b = -\nabla_b\xi_a$
and \cite{Wald1984},
\begin{align} \label{KillingRiemann}
  \nabla_a\nabla_b\xi_c = R_{bcad}\xi^d.
\end{align}
If the Killing vector is hypersurface orthogonal, then it satisfies \cite{Wald1984}
\begin{align} \label{HypersurfaceOrthogonal}
  \xi_{[a}\nabla_b\xi_{c]} = 0.
\end{align}
From this we get the useful expression
\begin{align} \label{Useful1}
  \xi_{[a} \nabla_{b]}\xi_c = \frac{1}{2} \xi_c \nabla_b \xi_a.
\end{align}
For hypersurface orthogonal Killing vectors, the Ricci tensor
satisfies $\xi_{[a}R_{a]b}\xi^b=0$  \cite{Carter1969},
\cite{Lazkoz2003}, and via Einstein's equations the
energy-momentum tensor satisfies
\begin{align} \label{EnMoHypOrt}
  T_{ab}\xi^b &= \alpha\xi_a,  
\end{align}
where
\begin{align}
  \pounds_\xi \alpha &= \xi^a\nabla_a  \alpha = 0.
\end{align}
When there are more than one matter field present this, in general, only
applies to the total energy-momentum tensor. Note also that the result
does not apply to test fields.

\section{Einstein-Klein-Gordon theory}
In this section we prove that in the Einstein-Klein-Gordon theory,
with a possible cosmological constant, the
superenergy current of the scalar
(Klein-Gordon) field for a hypersurface orthogonal Killing vector is
proportional to the Killing vector. The energy-momentum tensor is
given by 
\begin{align}
  T_{ab} = -\nabla_a\phi\nabla_b\phi +
  \frac{1}{2}g_{ab}( \nabla_c\phi\nabla^c\phi + m^2\phi^2),
\end{align}
where the scalar field, $\phi$, satisfies the Klein-Gordon equation,
$\nabla^c\nabla_c\phi = m^2\phi$. The superenergy tensor of the scalar field
is given by \cite{Senovilla2000}
\begin{align}
  S_{abcd}  
=& \nabla_a\nabla_c\phi\nabla_b\nabla_d\phi + 
             \nabla_a\nabla_d\phi\nabla_b\nabla_c\phi
  - g_{ab}( \nabla_c\nabla^e\phi\nabla_d\nabla_e\phi + m^2
  \nabla_c\phi\nabla_d\phi )\nonumber\\
 &-g_{cd}( \nabla_a\nabla^e\phi\nabla_b\nabla_e\phi + m^2
  \nabla_a\phi\nabla_b\phi )\nonumber\\
 &+\frac{1}{2}g_{ab}g_{cd}( \nabla_e\nabla_f\phi\nabla^e\nabla^f\phi
   + 2m^2\nabla_e\phi\nabla^e\phi + m^4\phi^2).
\end{align}
It has the following symmetries, $S_{abcd} = S_{(ab)(cd)} = S_{cdab}$.
Contracting thrice with $\xi^a$ gives the current
\begin{align} \label{STemp1}
  S_{abcd}\xi^b\xi^c\xi^d = &2
  \nabla_a\nabla_c\phi\nabla_b\nabla_d\phi\xi^b\xi^c\xi^d \nonumber\\
&- 
  \xi^c\xi_c\xi^b( \nabla_a\nabla^e\phi\nabla_b\nabla_e\phi + m^2
  \nabla_a\phi\nabla_b\phi ) + \omega\xi_a,
\end{align}
where $\omega$ is used to collect the proportionality factors of $\xi_a$.
We will split this into two cases depending on whether the scalar
field is massive ($m\neq 0$) or massless ($m=0$). In the first case
the Lie-derivative of the field vanishes for Killing vectors,
$\xi^a\nabla_a \phi = 0$ and in the second case it is constant,
$\xi^a\nabla_a\phi = {\rm C}$ \cite{Senovilla2000}.

We start with the massive case. Using the Leibniz rule on the first term
of (\ref{STemp1}), taking exterior product with $\xi_e$ and using
(\ref{Useful1}) gives
\begin{align}
  \xi_{[e}\nabla_{a]}\nabla_c\phi\nabla_b\nabla_d\phi \xi^b\xi^c\xi^d
  = &
  \xi_{[e}\nabla_{]a}(\xi^c\nabla_c\phi)\nabla_b\nabla_d\phi\xi^b\xi^d
 -\xi_{[e}\nabla_{a]}\xi^c\nabla_c\phi\nabla_b\nabla_d\phi\xi^b\xi^d
 \nonumber\\
 =&
 -\frac{1}{2}\xi^c\nabla_a\xi_e\nabla_c\phi\nabla_b\nabla_d\phi\xi^b\xi^d
 = 0.
\end{align}
Multiplying the second term of (\ref{STemp1}) with $\xi_e$ and using the
Leibniz rule and (\ref{HypersurfaceOrthogonal}) gives
\begin{align}
  \xi_e \nabla_a\nabla_c\phi\nabla^c\nabla_b\phi \xi^b = &
   - \xi_e\nabla_a\nabla_c\phi\nabla^c\xi^b\nabla_b\phi =
   \nabla_a\nabla_c\phi \xi^c\nabla^b\xi_e\nabla_b\phi
  =\nabla^c\xi_a\nabla_c\phi\nabla^b\xi_e\nabla_b\phi.
\end{align}
When antisymmetrizing over the indices $ea$ this expression then vanishes. Hence,
\begin{align} \label{STemp2}
  \xi_{[e}S_{a]bcd}\xi^b\xi^c\xi^d = 0.
\end{align}

For the massless case we note that (\ref{EnMoHypOrt}) implies that
$\nabla_a\phi = \alpha'\xi_a$. The first term of (\ref{STemp1}) then
equals
\begin{align}
  \nabla_a\nabla_c\phi\nabla_b\nabla_d\phi\xi^b\xi^c\xi^d
 =-\nabla_a\nabla_c\phi\nabla_b\xi^d\nabla_d\phi\xi^b\xi^c
 =\nabla_a\nabla_c\phi\nabla_b\xi^d\alpha'\xi_d\xi^b\xi^c=0.
\end{align}
The second term of (\ref{STemp1}) is treated in the same way as the
massive case and (\ref{STemp2}) thus holds in this case as well. 

\begin{theorem}
  If $\xi^a$ is a hypersurface orthogonal Killing vector, then the
  superenergy tensor of the scalar field in Einstein-Klein-Gordon
  theory, possibly with a cosmological constant, $\Lambda$, is
  proportional to $\xi^a$ and conserved,
\begin{align}
  S_{abcd}\xi^b\xi^c\xi^d &= \gamma\xi_a, &
  \nabla^a ( S_{abcd}\xi^b\xi^c\xi^d )&= 0.
\end{align}
\end{theorem}

The conservation of this current follows from the fact that the Bel
tensor together with the superenergy tensor of the scalar field gives
conserved currents for Killing vectors, and that, when the Killing
vectors are hypersurface orthogonal, the Bel currents are
independently conserved. 
Alternatively, observing that for Killing vectors,
$\pounds_\xi \nabla_a\phi
= \nabla_a\pounds_\xi\phi = 0$, the superenergy tensor will have
vanishing Lie-derivative with respect to $\xi^a$.
Then,
\begin{align}
  \nabla^a(S_{abcd}\xi^b\xi^c\xi^d) = \xi^a\nabla_a\gamma =
  \pounds_\xi \gamma =0 .
\end{align}
Note that this result implies that for a hypersurface orthogonal Killing
vector, the mixed superenergy current will be proportional to the
Killing vector,
\begin{align}
  (B_{abcd}+S_{abcd})\xi^b\xi^c\xi^d = \beta\xi_a.
\end{align}

\section{Einstein-Maxwell theory}
In this section we will prove that for a hypersurface orthogonal Killing
vector in Einstein-Maxwell theory, with a possible cosmological
constant, $\Lambda$, the corresponding Chevreton
current will be independently conserved and proportional to the
Killing vector. This is proven for the case when the electromagnetic
field is source-free and inherits the symmetry of the spacetime.
As a consequence of Lemma \ref{Lemma1} we also have
that a similar result holds for the trace of the Chevreton tensor.

The electromagnetic field is described by the Maxwell tensor, $F_{ab}
= - F_{ba}$, 
which in source-free regions satisfies
\begin{align}
  \nabla^a F_{ab} &= 0,  & \nabla_{[a}F_{cd]} = 0.
\end{align}
The energy momentum tensor is given by
\begin{align} \label{EMEnergyMomentum}
  T_{ab} = -F_{ac}F_b{}^c + \frac{1}{4} g_{ab}F_{cd}F^{cd}.
\end{align}
The Ricci scalar, $R$, satisfies $R = 4\Lambda$, where $\Lambda$
is the cosmological constant.
From (\ref{EnMoHypOrt}) we then have that
\begin{align} \label{EnergyMomentumHypersurface}
  T_{ab}\xi^b =  ( - F_{ac} F_b{}^c + \frac{1}{4}g_{ab}F_{cd}F^{cd} )
  \xi^b = \alpha\xi_a,
\end{align}
or
\begin{align}
  F_{ac}F_b{}^c \xi^b = \alpha' \xi_a.
\end{align}
Generally, the Lie-derivative of the Maxwell field in four-dimensional
Einstein-Maxwell
theory satisfies
\cite{Michalski1975}, \cite{Stephani2003},
\begin{align}
  \pounds_\xi F_{ab} = \xi^c\nabla_c F_{ab} + F_{cb}\nabla_a\xi^c +
  F_{ac}\nabla_b\xi^c = k \stackrel{*}F_{ab},
\end{align}
where $k$ is a constant and $\stackrel{*}F_{ab}$ is the Hodge dual of
$F_{ab}$. We will assume that the electromagnetic field
inherits (or is admitted by) the symmetry of the spacetime. Then $k=0$
and we have the useful rearrangement,
\begin{align} \label{MaxwellLieDerivative}
  \xi^c\nabla_c F_{ab} = - F_{cb}\nabla_a\xi^c - F_{ac}\nabla_b\xi^c.
\end{align}
The basic superenergy tensor of the electromagnetic field is given by \cite{Senovilla2000},
\begin{align} \label{BasicSE}
  E_{abcd} = &-\nabla_a F_{ce} \nabla_b F_d{}^e 
             -\nabla_b F_{ce} \nabla_a F_d{}^e 
             +g_{ab}\nabla_f F_{ce} \nabla^f F_d{}^e \nonumber\\
             &+\frac{1}{2} g_{cd}\nabla_a F_{ef} \nabla_b F^{ef}
   -\frac{1}{4} g_{ab}g_{cd}\nabla_e F_{fg} \nabla^e F^{fg}.
\end{align}
This tensor is symmetric in the first index pair and in the second
index pair, $E_{abcd}=E_{(ab)(cd)}$. The Chevreton tensor is then
defined as $H_{abcd} = \frac{1}{2}(E_{abcd}+E_{cdab})$, or
\begin{align} \label{Chevreton}
     H_{abcd} = 
  & -\frac{1}{2}( \nabla_a F_{ce} \nabla_b F_d{}^e + 
                             \nabla_b F_{ce} \nabla_a F_d{}^e +
                             \nabla_c F_{ae} \nabla_d F_b{}^e +
                             \nabla_d F_{ae} \nabla_c F_b{}^e ) \nonumber\\
             & +\frac{1}{2}( g_{ab}\nabla_f F_{ce} \nabla^f F_d{}^e +
                             g_{cd}\nabla_f F_{ae} \nabla^f F_b{}^e)
              +\frac{1}{4}( g_{ab}\nabla_c F_{ef} \nabla_d F^{ef} +
                             g_{cd}\nabla_a F_{ef} \nabla_b F^{ef})\nonumber\\
             & -\frac{1}{4} g_{ab}g_{cd}\nabla_e F_{fg} \nabla^e F^{fg}.
\end{align}
This tensor has the same obvious symmetries as the Bel tensor, $H_{abcd}
= H_{(ab)(cd)} = H_{cdab}$, but it is actually completely symmetric in
four dimensions, $H_{abcd} = H_{(abcd)}$, as was shown in \cite{Bergqvist2003}.
This tensor is 
more interesting physically than the basic superenergy tensor, because it gives unique
currents and a unique divergence.
Contracting thrice with $\xi^a$ gives us the following current 
\begin{align} \label{ChevretonCurrent1}
     H_{abcd}\xi^b\xi^c\xi^d = & 
   - ( \nabla_a F_{ce} \nabla_b F_d{}^e + 
       \nabla_c F_{ae} \nabla_d F_b{}^e )\xi^b\xi^c\xi^d \nonumber\\
 & + \frac{1}{4}\xi^c\xi_c( 2\nabla_f F_{ae} \nabla^f F_b{}^e + 
                             \nabla_a F_{ef} \nabla_b F^{ef}) \xi^b +
                             \omega\xi_a,
\end{align}
where $\omega$ again is used to collect the proportionality factors of $\xi^a$. We would
like to show that the remaining terms are also proportional to
$\xi^a$. We split the problem into three parts. We start by examining
the third term of (\ref{ChevretonCurrent1}) followed by the second
term and then finally the first and fourth terms together.

\begin{lemma} \label{Lemma1}
Under our assumptions,
\begin{align} \label{Lemma1eq}
  \xi_{[g} \nabla_{|f|} F_{a]e} \nabla^f F_b{}^e \xi^b = 0.
\end{align}
\end{lemma}

\begin{proof}
Applying $\nabla^f\nabla_f$ to the energy-momentum tensor (\ref{EMEnergyMomentum}) gives us
\begin{align} \label{Temp2}
  2 \nabla_f F_{ae} \nabla^f F_b{}^e \xi^b = -\xi^b \nabla^f\nabla_f
  T_{ab} - \xi^bF_{ae}\nabla^f\nabla_f F_b{}^e - \xi^b F_b{}^e
  \nabla^f\nabla_f F_{ae} + \omega\xi_a.
\end{align}
The first term of the right hand side is rewritten with the Leibniz rule
as
\begin{align} \label{Temp1}
 \xi^b\nabla^f\nabla_f T_{ab}
=&  \nabla^f( \xi^b\nabla_f T_{ab}) - \nabla^f\xi^b \nabla_f T_{ab}
  =\nabla^f( \nabla_f( \xi^bT_{ab} ) -T_{ab}\nabla_f\xi^b) -
  \nabla^f\xi^b \nabla_f T_{ab} \nonumber\\
=&
  \nabla^f\nabla_f( \alpha \xi_a) - 2\nabla^f\xi^b \nabla_f T_{ab} -
  T_{ab}\nabla^f\nabla_f \xi^b,
\end{align}
where in the last step (\ref{EnergyMomentumHypersurface}) was used.
Here,
\begin{align}
  \nabla^f\nabla_f( \alpha\xi_a ) = \alpha\nabla^f\nabla_f \xi_a +
  2\nabla^f\xi_a \nabla_f \alpha + \xi_a\nabla^f\nabla_f\alpha.
\end{align}
We note that, by (\ref{KillingRiemann}),
\begin{align} \label{WaveKilling}
  \nabla^b\nabla_b \xi_a = R_{ba}{}^b{}_c\xi^c =  R_{ac}\xi^c
  = -T_{ac}\xi^c =- \alpha\xi_a.
\end{align}
Hence, with (\ref{Useful1}), 
\begin{align}
  \nabla^f\nabla_f( \alpha\xi_{[a} )\xi_{e]} =
  -2\xi_{[e}\nabla^f\xi_{a]}\nabla_f\alpha
  = 2 \xi_{[e}\nabla_{a]}\xi^f\nabla_f\alpha =  \nabla_a\xi_e
  \xi^f\nabla_f\alpha = 0.
\end{align}
The third term in the last expression of (\ref{Temp1}) gives
\begin{align}
  \xi_{[e}T_{a]b}\nabla^f\nabla_f \xi^b = - \alpha \xi_{[e}
  T_{a]b}\xi^b =  - \alpha^2\xi_{[e}\xi_{a]} = 0.
\end{align}
To simplify the second term of expression (\ref{Temp1}) takes somewhat more effort. We start by looking at
\begin{align}
  \xi_e \nabla^f\xi^b\nabla_f T_{ab} = - \xi^f \nabla^b\xi_e \nabla_f
  T_{ab} - \xi^b \nabla_e\xi^f \nabla_f T_{ab},
\end{align}
by (\ref{HypersurfaceOrthogonal}). The energy-momentum tensor has
vanishing Lie-derivative with respect to $\xi^a$, $\pounds_\xi T_{ab} = 0$,
giving us
\begin{align} \label{Temp4}
  \xi^c\nabla_c T_{ab} = - T_{cb}\nabla_a\xi^c - T_{ac}\nabla_b\xi^c.
\end{align}
This together with the Leibniz rule then yields
\begin{align}
 \xi_e \nabla^f\xi^b\nabla_f T_{ab} = &\nabla^b\xi_e( T_{fb}\nabla_a\xi^f + T_{af}\nabla_b\xi^f)
  -\nabla_e\xi^f \nabla_f( T_{ab}\xi^b ) +\nabla_e\xi^f
  T_{ab}\nabla_f\xi^b \nonumber\\
 =& - T_{fb}\nabla^b\xi_e\nabla^f\xi_a + \alpha\nabla^f \xi_e\nabla_f
 \xi_a
   - \xi_a \nabla_e\xi^f\nabla_f\alpha.
\end{align}
The first two terms are symmetric in $e$ and $a$, so
antisymmetrization gives, with (\ref{Useful1}),
\begin{align}
  \xi_{[e} \nabla^f\xi^b\nabla_{|f|} T_{a]b} =
  -\xi_{[a}\nabla_{e]}\xi^f\nabla_f\alpha
  = - \frac{1}{2} \nabla_e\xi_a \xi^f\nabla_f\alpha = 0.
\end{align}
Hence, 
\begin{align} \label{Temp15}
  \xi_{[e}\nabla^f\nabla_{|f|} T_{a]b}\xi^b = 0.
\end{align}
We now continue with the second and third terms on the right hand side
of (\ref{Temp2}).
The Maxwell wave equation in four dimensions for source-free regions
is given by \cite{Andersson1996}
\begin{align} \label{MWE}
  \nabla^f\nabla_f F_{ab} = 2C^c{}_{ab}{}^d F_{dc} - \frac{1}{3}R F_{ab}.
\end{align}
The second term of (\ref{Temp2}) can therefore be rewritten as
\begin{align}
  \xi^b F_{ae}\nabla^f\nabla_f F_b{}^e = 2\xi^b
  F_{ae}F_{dc}C^c{}_b{}^{ed}  - \frac{1}{3}R\xi^b F_{ae}F_b{}^e.
\end{align}
The second and third terms of (\ref{Temp2}) together then equal
\begin{align} \label{Temp16}
 & 2\xi^b( F_a{}^e F^{dc} C_{cbed} + F_b{}^e F^{dc} C_{caed} ) -
  \frac{2}{3}R\alpha'\xi_a.
\end{align}
We attack this by using a dimensionally dependent identity \cite{Lovelock1970},
\begin{align}
  C_{[ab}{}^{[cd} \delta_{e]}^{f]} = 0,
\end{align}
which holds in (and only in) four dimensions. Contracting this with $F^a{}_dF^e{}_c$
yields
\begin{align}
&  C_{ab}{}^{cd}F^a{}_dF^f{}_c + C_{ab}{}^{fc}F^a{}_eF^e{}_c
 +C_{be}{}^{cd}F^f{}_dF^e{}_c + C_{be}{}^{df}F^c{}_dF^e{}_c \nonumber\\
&
 +C_{ea}{}^{cd}F^a{}_dF^e{}_c\delta_b^f + C_{ea}{}^{df}F^a{}_dF^e{}_b
 +C_{ea}{}^{fc}F^a{}_bF^e{}_c = 0.
\end{align}
Identifying terms 1 and 3, 2 and 4, and 6 and 7, respectively,
gives us
\begin{align}
  C_{ea}{}^{df}F^a{}_dF^e{}_b + C_{ab}{}^{cd}F^a{}_dF^f{}_c
 +C_{ab}{}^{fc}F^a{}_eF^e{}_c +
 \frac{1}{2}C_{ea}{}^{cd}F^a{}_dF^e{}_c\delta_b^f =0.
\end{align}
Thus,
\begin{align} \label{Temp14}
 \xi^b( F_a{}^eF^{dc} C_{cbed} + F_b{}^eF^{dc} C_{caed} ) =
C_{cbad}F^{ce}F_e{}^d\xi^b + \frac{1}{2}C_{bcde}F^{bd}F^{ce}\xi_a.
\end{align}
Rewriting the Weyl tensor in terms of the Riemann tensor and using
(\ref{EnergyMomentumHypersurface}) then gives
\begin{align}
  C_{cbad}F^{ce}F_e{}^d\xi^b = R_{cbad}F^{ce}F_e{}^d\xi^b + \omega\xi_a.
\end{align}
From (\ref{KillingRiemann}) we have that
\begin{align}
   R_{cbad}F^{ce}F_e{}^d\xi^b = F^{ce}F_e{}^d\nabla_c\nabla_a\xi_d.
\end{align}
Taking the covariant derivative of (\ref{HypersurfaceOrthogonal}) yields
\begin{align}
  \nabla_c ( \xi_{[b}\nabla_a\xi_{d]} ) = 0,
\end{align}
or
\begin{align}
 2 \xi_{[b} \nabla_{|c|}\nabla_{a]}\xi_d =
  -\nabla_c\xi_b\nabla_a\xi_d - \nabla_c\xi_a\nabla_d\xi_b
  -\nabla_c\xi_d\nabla_b\xi_a - \xi_d\nabla_c\nabla_b\xi_a.
\end{align}
This gives us that
\begin{align}
  2 \xi_{[b} F^{ce}F_{|e}{}^d\nabla_{c|}\nabla_{a]}\xi_d =
   -\xi_d F^{ce}F_e{}^d\nabla_c\nabla_b\xi_a=  \alpha' \xi^c R_{bace}\xi^e = 0.
\end{align}
Thus, taking exterior product of (\ref{Temp14}) with $\xi_g$ yields zero and
this together with (\ref{Temp15}) gives us that
\begin{align}
  \xi_{[g} \nabla_{|f|} F_{a]e} \nabla^f F_b{}^e \xi^b = 0.
\end{align}
\end{proof}
We have used four dimensions explicitly in this proof. If one looks at
arbitrary dimension $n$, for inherited symmetry,
(\ref{MaxwellLieDerivative}) is still valid, but the Maxwell wave equation (\ref{MWE}) is
replaced by \cite{Andersson1996}
\begin{align}
  \nabla^f\nabla_f F_{ab} = 2C^c{}_{ab}{}^d F_{dc} -
  \frac{2(n-2)}{n(n-1)}R F_{ab} + 2\frac{n-4}{n-2}F_{c[a}T_{b]}{}^c.
\end{align}
In the calculations above the last term here also gives terms
proportional to $\xi_a$. However, it seems that in (\ref{Temp16}) we
still need to restrict to four dimensions, but this is an open question.

We note here that Lemma \ref{Lemma1} can be applied to the
trace of the Chevreton tensor, which is given by \cite{Bergqvist2003}
\begin{align}
  H_{ab} = H_{abc}{}^c = \nabla_c F_{ad}\nabla^c F_b{}^d 
   - \frac{1}{4} g_{ab} \nabla_c F_{de}\nabla^c F^{de}.
\end{align}
Contracting with $\xi^b$ and taking exterior product with $\xi_e$ and
using Lemma \ref{Lemma1} then implies that
\begin{align}
  \xi_{[e} H_{a]b}\xi^b = \xi_{[e} \nabla_{|c|} F_{a]d}\nabla^c
  F_b{}^d\xi^b = 0.
\end{align}
Therefore, $H_{ab}\xi^b = \gamma\xi_a$. We know from \cite{Bergqvist2003}
that $H_{ab}$ is symmetric, trace-free, and divergence-free in
four-dimensional Einstein-Maxwell theory with a source-free
electromagnetic field, so this current is divergence-free.

\begin{theorem}
Assume that we have four-dimensional Einstein-Maxwell theory, possibly with a
cosmological constant $\Lambda$, with a source-free
electromagnetic field that inherits the symmetry of the spacetime.
If $\xi^a$ is a hypersurface orthogonal Killing
vector, then the current $H_{ab}\xi^b$, where $H_{ab}$ is the trace
of the Chevreton tensor, is proportional to $\xi_a$ and divergence-free,
\begin{align}
  H_{ab}\xi^b &= \gamma\xi_a, & \nabla^a( H_{ab}\xi^b ) &= 0.
\end{align}
\end{theorem}

Going back now to the Chevreton current (\ref{ChevretonCurrent1}), we
turn our attention to the second term on the right-hand side.
\begin{lemma} \label{Lemma2}
Under our assumptions,
\begin{align} \label{Lemma2eq}
  \xi_{[f} \xi^b\xi^c\xi^d \nabla_{|c|}F_{a]e}\nabla_d F_b{}^e = 0.
\end{align}
\end{lemma}

\begin{proof}
Using (\ref{MaxwellLieDerivative}) we have that
\begin{align}
  \xi^b\xi^c\xi^d \nabla_c F_{ae}\nabla_d F_b{}^e= &
   \xi^b( F_{ce}\nabla_a\xi^c+ F_{ac}\nabla_e\xi^c)
        ( F_d{}^e\nabla_b\xi^d+ F_{bd}\nabla^e\xi^d) \nonumber\\
= & \xi^b F_{ce}F_d{}^e\nabla_a\xi^c \nabla_b\xi^d
  + \xi^b F_{ce}F_{bd}\nabla_a\xi^c \nabla^e\xi^d \nonumber\\
 &+ \xi^b F_{ac}F_d{}^e\nabla_e\xi^c \nabla_b\xi^d
  + \xi^b F_{ac}F_{bd}\nabla_e\xi^c \nabla^e\xi^d.
\end{align}
Here then, using (\ref{Useful1}), the two first terms give
\begin{align}
  \xi_{[f} \nabla_{a]}\xi^c \xi^b( F_{ce} F_d{}^e \nabla_b\xi^d +
  F_{ce}F_{bd} \nabla^e\xi^d) =& \frac{1}{2}\nabla_a\xi_f
  \xi^c\xi^b( F_{ce} F_d{}^e \nabla_b\xi^d +
   F_{ce}F_{bd} \nabla^e\xi^d) \nonumber\\
 =& \frac{1}{2} \nabla_a\xi_f \alpha'\xi^b \xi_d \nabla_b\xi^d = 0.
\end{align}
The two last terms, by (\ref{Useful1}), equal
\begin{align}
  F_{ac}F_{de}\nabla^b\xi^d( \xi_b\nabla^e\xi^c - \xi^e\nabla_b\xi^c )
 = - F_{ac}F_{de}\xi^c\nabla^b\xi^d  \nabla_b\xi^e = 0.
\end{align}
Hence,
\begin{align} 
  \xi_{[f} \xi^b\xi^c\xi^d \nabla_{|c|}F_{a]e}\nabla_d F_b{}^e = 0.
\end{align}
\end{proof}

The remaining two terms of (\ref{ChevretonCurrent1}), the first and
the fourth, are treated together.

\begin{lemma} \label{Lemma3}
Under our assumptions,
\begin{align} \label{Lemma3eq}
- \xi_{[g}  \nabla_{a]} F_{ce}\nabla_b F_{d}{}^e\xi^b\xi^c\xi^d
   + \frac{1}{4} \xi_{[g}\xi^c\xi_{|c|} \nabla_{a]} F_{ef}\nabla_b
   F^{ef}\xi^b = 0.
\end{align}
\end{lemma}

\begin{proof}
Taking two derivatives of the energy-momentum tensor yields
\begin{align} \label{Temp3}
 &- \nabla_a F_{ce}\nabla_b F_d{}^e\xi^b\xi^c\xi^d
   + \frac{1}{4} \xi^c\xi_c \nabla_a F_{ef}\nabla_b F^{ef}\xi^b \nonumber\\
=& \frac{1}{2}\xi^b\xi^c\xi^d \nabla_a\nabla_b T_{cd}+ F_{ce}\nabla_a\nabla_b F_d{}^e\xi^b\xi^c\xi^d -
 \frac{1}{4}\xi^c\xi_c F_{ef}\nabla_a \nabla_b F^{ef}\xi^b 
   .
\end{align}
We start with the first term of the right-hand side and rewrite it
by the Leibniz rule,
\begin{align}
  \xi^b\xi^c\xi^d \nabla_a \nabla_b T_{cd} = &
   \xi^b\xi^c \nabla_a( \xi^d \nabla_b T_{cd} ) - \xi^b\xi^c \nabla_b
   T_{cd}\nabla_a\xi^d \nonumber\\
 =&
   \xi^b\xi^c\nabla_a( \nabla_b( \xi^d T_{cd} ) - T_{cd}\nabla_b
   \xi^d)
   - \xi^b\nabla_b( \xi^c T_{cd} )\nabla_a\xi^d + \xi^b T_{cd}
   \nabla_b\xi^c \nabla_a \xi^d \nonumber\\
 =&
   \xi^b\xi^c\nabla_a\nabla_b( \xi^d T_{cd} ) 
   - \xi^b \nabla_a ( \xi^c T_{cd})\nabla_b \xi^d - \xi^b\xi^c
   T_{cd}\nabla_a\nabla_b\xi^d \nonumber\\
  &-\xi^b\nabla_b(\xi^c T_{cd})\nabla_a\xi^d +
  2 \xi^bT_{cd}\nabla_b\xi^c\nabla_a\xi^d.
\end{align}
Using (\ref{EnergyMomentumHypersurface}) gives
\begin{align} \label{Temp13}
 & \xi^b\xi^c \nabla_a\nabla_b( \alpha\xi_c) - \xi^b \nabla_a(
  \alpha\xi_d)\nabla_b\xi^d
 - \alpha\xi^b\xi_d \nabla_a\nabla_b\xi^d - \xi^b\nabla_b(
 \alpha\xi_d)\nabla_a\xi^d
 + 2\xi^b T_{cd}\nabla_b\xi^c\nabla_a\xi^d \nonumber\\
=& \xi^b\xi^c \nabla_a\nabla_b( \alpha\xi_c)
   - 2 \alpha\xi^b \nabla_a \xi_d \nabla_b\xi^d
   + 2\xi^b T_{cd}\nabla_b\xi^c\nabla_a\xi^d.
\end{align}
Expanding the first term and applying the Leibniz rule yields
\begin{align}
  \xi^b\xi^c\nabla_a\nabla_b( \alpha\xi_c ) = \xi^c\xi_c \xi^b
  \nabla_a\nabla_b\alpha
  = -\xi^c\xi_c \nabla_a\xi^b\nabla_b\alpha.
\end{align}
This then gives us, by (\ref{Useful1}),
\begin{align}
  \xi_{[e}\xi^b\xi^c\nabla_{a]}\nabla_b( \alpha\xi_c) =
  -\frac{1}{2}\xi^c\xi_c \nabla_a\xi_e \xi^b\nabla_b\alpha = 0.
\end{align}
Continuing with the second term of (\ref{Temp13}),
\begin{align}
   \xi_{[e}\nabla_{a]}\xi_d \alpha\xi^b \nabla_b\xi^d =
   \frac{1}{2}\alpha\nabla_a\xi_e \xi_d\xi^b \nabla_b\xi^d = 0.
\end{align}
The last term of (\ref{Temp13}) gives us
\begin{align}
  \xi_{[e}\nabla_{a]}\xi^d T_{cd}\xi^b\nabla_b\xi^c = 
  \frac{1}{2} \nabla_a\xi_e \alpha\xi_c\xi^b\nabla_b\xi^c = 0.
\end{align}
Hence,
\begin{align}
\xi_{[e}\xi^b\xi^c\xi^d \nabla_{a]} \nabla_b T_{cd} = 0.
\end{align}
The two last terms of (\ref{Temp3}) are rewritten by taking a covariant
derivative of (\ref{MaxwellLieDerivative}), i.e., expanding
$\nabla_a\pounds_\xi F_{bc}=0$, which gives us
\begin{align}
  F_{ce}\nabla_a\nabla_b F_d{}^e\xi^b\xi^c\xi^d = &
  -F_{ce}\xi^c\xi^d ( \nabla_b F_d{}^e \nabla_a \xi^b + \nabla_a
  F_b{}^e\nabla_d\xi^b \nonumber\\
 & + \nabla_a F_{db}\nabla^e\xi^b + F_b{}^eR_d{}^b{}_{af}\xi^f
   + F_{db}R^{eb}{}_{af}\xi^f)
\end{align}
and
\begin{align}
   F_{ef}\nabla_a\nabla_b F^{ef}\xi^b =&
   -F_{ef}( \nabla_a \xi^b \nabla_b F^{ef} + 2\nabla_a
   F_b{}^f\nabla^e\xi^b).
\end{align}
Both terms involving the Riemann tensor disappear because of (\ref{EnergyMomentumHypersurface})
and symmetric-antisymmetric contractions. Hence,
\begin{align} \label{Temp5}
 &  F_{ce}\nabla_a\nabla_b F_d{}^e\xi^b\xi^c\xi^d 
  -\frac{1}{4}\xi^c\xi_c  F_{ef}\nabla_a\nabla_b F^{ef}\xi^b  \nonumber\\
=& \nabla_a\xi^b( -\xi^c\xi^d F_{ce}\nabla_b F_d{}^e +
\frac{1}{4}\xi^c\xi_c F_{ef}\nabla_b F^{ef}) \nonumber\\
&
- \xi^c\xi^d F_{ce }\nabla_a F_b{}^e\nabla_d\xi^b 
  -\xi^c\xi^d F_{ce } \nabla_a F_{db}\nabla^e\xi^b
 +\frac{1}{2}\xi^c\xi_cF_{ef}\nabla_a F_b{}^f\nabla^e\xi^b.
\end{align}
We note that the two first terms on the right-hand side can be rewritten as
\begin{align}
 \nabla_a\xi^b(-\xi^c\xi^d F_{ce}\nabla_b F_d{}^e +
 \frac{1}{4}\xi^c\xi_c F_{ef}\nabla_b F^{ef}) =  \frac{1}{2} \xi^c\xi^d \nabla_a\xi^b \nabla_b T_{cd} .
\end{align}
Then, by use of (\ref{Useful1}), (\ref{Temp4}), and (\ref{EnergyMomentumHypersurface}),
\begin{align}
  \xi_{[e}\xi^c\xi^d\nabla_{a]}\xi^b \nabla_b T_{cd}
  =\frac{1}{2} \nabla_a\xi_e \xi^b\xi^c\xi^d \nabla_b T_{cd}
  = -\nabla_a\xi_e \xi^c\xi^d T_{bd}\nabla_c\xi^b
  = -\nabla_a\xi_e \alpha\xi^c\xi_b\nabla_c\xi^b = 0.
\end{align}
The last term of (\ref{Temp5}) is rewritten using (\ref{Useful1}) and
the Leibniz rule as
\begin{align}
  \frac{1}{2}\xi^c\xi_c F_{ef}\nabla_aF_b{}^f \nabla^e\xi^b =&
  -\frac{1}{2}\xi^e\xi_c F_{ef}\nabla_a F_b{}^f \nabla^b\xi^c 
 + \frac{1}{2}\xi^b\xi_c F_{ef}\nabla_a F_b{}^f \nabla^e\xi^c
 \nonumber\\
=&
  -\frac{1}{2}\xi^e\xi_c\nabla_a(F_{ef}F_b{}^f)\nabla^b\xi^c + \xi^b\xi_c
  F_{ef}\nabla_a F_b{}^f \nabla^e\xi^c.
\end{align}
Hence, the third and last terms of (\ref{Temp5}), using the Leibniz rule
and (\ref{EnergyMomentumHypersurface}), they equal
\begin{align}
  &-\xi^c\xi^d F_{ce}\nabla_a F_b{}^e \nabla_d \xi^b + \xi^b\xi_c
  F_{ef}\nabla_a F_b{}^f \nabla^e \xi^c  -
  \frac{1}{2}\xi^e\xi_c\nabla_a( F_{ef} F_b{}^f )\nabla^b\xi^c \nonumber\\
=&
 -  \frac{1}{2}  \xi^c\xi^d\nabla_d\xi^b \nabla_a( F_{ce}F_b{}^e)
= -\frac{1}{2} \alpha' \xi^d\nabla_d\xi^b \nabla_a \xi_b  
+ \frac{1}{2} \xi^d\nabla_d\xi^b F_{ce}F_b{}^e \nabla_a\xi^c.
\end{align}
Taking exterior product with $\xi_f$ and using (\ref{Useful1}) and
(\ref{EnergyMomentumHypersurface}) then
yields
\begin{align}
  -\frac{1}{4}\alpha' \xi^d\nabla_d\xi^b \xi_b \nabla_a\xi_f 
 + \frac{1}{4} \xi^d\nabla_d\xi^b F_{ce}F_b{}^e \xi^c\nabla_a\xi_f
 = 0.
\end{align}
The fourth term of (\ref{Temp5}) is rewritten using (\ref{Useful1}) as
\begin{align}
  - \xi^c\xi^d F_{ce} \nabla_a F_{db} \nabla^e\xi^b =
  -\xi^d F_{ce}\nabla_a F_{db} \xi^{[c}\nabla^{e]}\xi^b =
  -\frac{1}{2} \xi^d F_{ce} \nabla_a F_{db} \xi^b \nabla^e\xi^c = 0.
\end{align}
Thus, taken together, we have that the exterior product of $\xi_g$ and
(\ref{Temp3}) equals
\begin{align}
- \xi_{[g}  \nabla_{a]} F_{ce}\nabla_b F_{d}{}^e\xi^b\xi^c\xi^d
   + \frac{1}{4} \xi_{[g}\xi^c\xi_{|c|} \nabla_{a]} F_{ef}\nabla_b
   F^{ef}\xi^b = 0.
\end{align}
\end{proof}

By (\ref{Chevreton}) and Lemma \ref{Lemma1}, Lemma \ref{Lemma2}, and
Lemma \ref{Lemma3}, we have
that
\begin{align}
  \xi_{[e} H_{a]bcd}\xi^b\xi^c\xi^d = 0
\end{align}
and therefore
\begin{align}
  H_{abcd}\xi^b\xi^c\xi^d = \gamma\xi_a.
\end{align}
The Lie-derivative commutes with the covariant derivative for
Killing vectors and hence, for inherited symmetry,
\begin{align}
  \pounds_\xi \nabla_a F_{bc} = \nabla_a \pounds_\xi F_{bc} =0.
\end{align}
This implies that the Chevreton tensor has vanishing Lie-derivative
with respect to Killing vectors,
\begin{align}
  \pounds_\xi H_{abcd} = 0.
\end{align}
Hence,
\begin{align}
  \pounds_\xi( H_{abcd}\xi^b\xi^c\xi^d ) = \pounds_\xi(\gamma\xi_a) = 
  \xi_a \xi^b\nabla_b \gamma = 0
\end{align}
and therefore
\begin{align}
  \nabla^a( H_{abcd}\xi^b\xi^c\xi^d ) = \nabla^a( \gamma\xi_a ) 
  = \gamma\nabla^a\xi_a + \xi_a \nabla^a \gamma = 0.
\end{align}
Thus, we have the following result

\begin{theorem} \label{Theorem2}
Assume that we have four-dimensional Einstein-Maxwell theory, possibly with a
cosmological constant, $\Lambda$, with a source-free 
electromagnetic field that inherits the symmetry of the spacetime.
If $\xi^a$ is a hypersurface orthogonal Killing vector, then the Chevreton current $H_{abcd}\xi^b\xi^c\xi^d$, is proportional to $\xi^a$ and divergence-free,
\begin{align}
  H_{abcd}\xi^b\xi^c\xi^d = \gamma\xi_a, & & \nabla^a(
  H_{abcd}\xi^b\xi^c\xi^d) =0.
\end{align}
\end{theorem}

Note that for the basic superenergy tensor, $E_{abcd}$, the two
possible currents, $E_{abcd}\xi^b\xi^c\xi^d$ and
$E_{cdab}\xi^b\xi^c\xi^d$, are both proportional to $\xi^a$. The first
one follows from Lemma \ref{Lemma3} and the second from Lemma
\ref{Lemma1} and Lemma \ref{Lemma2}. 
Since all terms in $E_{abcd}$ have vanishing Lie-derivative
with respect to $\xi^a$, they are all independently conserved. 
Hence, we can state the slightly more general result,

\begin{theorem} Under the conditions of Theorem \ref{Theorem2}, the
  two possible currents constructed from the basic superenergy tensor
  (\ref{BasicSE}) , $E_{abcd}\xi^b\xi^c\xi^d$ and
$E_{cdab}\xi^b\xi^c\xi^d$, are proportional to $\xi^a$ and
independently conserved,
\begin{align}
  E_{abcd}\xi^b\xi^c\xi^d = \gamma_1\xi_a, & & \nabla^a(
  E_{abcd}\xi^b\xi^c\xi^d) =0, \nonumber\\
  E_{cdab}\xi^b\xi^c\xi^d = \gamma_2\xi_a, & & \nabla^a(
  E_{cdab}\xi^b\xi^c\xi^d) =0.
\end{align}
\end{theorem}

These results can be applied to the timelike Killing
vector in static Einstein-Maxwell spacetimes such as Reissner-Nordström.

It can be shown that for four-dimensional Einstein-Maxwell theory, the Bel tensor is decomposed as
\begin{align}
  B_{abcd} = T_{abcd} + T_{ab}T_{cd} + \frac{1}{48} R^2g_{ab}g_{cd},
\end{align}
where $T_{abcd}$ is the Bel-Robinson tensor (\ref{BRTensor}).
The decomposition without a cosmological constant was given in \cite{Bonilla1997}.
Contracting thrice with $\xi^a$ and using (\ref{Lazkoz:eq}) and
(\ref{EnMoHypOrt}) yields
\begin{align}
  \gamma\xi_a = T_{abcd}\xi^b\xi^c\xi^d + \alpha^2\xi_a\xi^b\xi_b + \frac{1}{48}R^2\xi_a\xi^b\xi_b,
\end{align}
or
\begin{align}
  T_{abcd}\xi^b\xi^c\xi^d = \beta\xi_a.
\end{align}
All components of the Bel tensor have vanishing Lie-derivative with
respect to Killing vectors, so for hypersurface
orthogonal Killing vectors in Einstein-Maxwell theory, the Bel tensor
splits into independently conserved components, each proportional
to $\xi^a$.

\section{Example for non-inherited symmetry}

In \cite{McIntosh1978} McIntosh gives an example of an
Einstein-Maxwell spacetime that contains a hypersurface orthogonal
Killing vector for which the electromagnetic field does not inherit
the symmetry. This seems
to be the only such known example. The metric is given by
\begin{align}
  {\rm d}s^2 = -( {\rm d}t - b r^2{\rm d}\phi)^2
   + r^2{\rm d}\phi^2
   + \exp(b^2r^2)  ( {\rm d}z^2 + {\rm d}r^2).
\end{align}
The electromagnetic potential is given by
\begin{align}
  A_a = - \sqrt{2} \sin(2bz+C)( \delta_{ta} -br^2\delta_{\phi a}),
\end{align}
where $b$ and $C$ are constants. The spacetime has three Killing
vectors,
\begin{align}
  \xi_{1a} &= \delta_{ta}, & \xi_{2a} &= \delta_{\phi a}, & \xi_{3a} = \delta_{za},
\end{align}
of which the last the electromagnetic field does not inherit the symmetry,
\begin{align}
  \pounds_{\xi_3} F_{ab} \neq 0.
\end{align}
However, it is still the case here that the Chevreton tensor has
vanishing Lie-derivative,
\begin{align}
  \pounds_{\xi_3} H_{abcd} = 0,
\end{align}
and also that the Chevreton current associated with this Killing
vector is proportional to it,
\begin{align}
   H_{abcd}\xi_3^b\xi_3^c\xi_3^d = \xi_{3a}\frac{5b^4(3-b^2r^2)}{6},
\end{align}
so in this case we have the same result as in Theorem \ref{Theorem2}.

\section{Conclusion}

We have seen that for a hypersurface orthogonal Killing vector, both
the gravitational superenergy tensor and the matter superenergy
tensors give rise to independently conserved currents that are
proportional to the Killing vector, thus giving rise to conserved
quantities in these cases. For the Einstein-Maxwell case we have not
been able to prove an $n$-dimensional result. The results for the Bel
current and the superenergy current for the scalar field are valid in
$n$ dimensions, so it would be desirable to show this for
the Chevreton current as well. Our result is only shown for
electromagnetic fields that inherit the symmetry of the spacetime and
it would be interesting to see if the result holds in the non-inheriting case
as well. The example given above gives support to this case.

For the scalar field in Einstein-Klein-Gordon theory, it has previously
been proven that when a Killing vector is present, the Bel tensor
together with the superenergy tensor of the
scalar field gives rise to a conserved mixed current governing
interchange of superenergy between the gravitational field and the
scalar field. 

The question of whether a similar construction is possible for
Einstein-Maxwell theory is still open. The result presented here,
Theorem \ref{Theorem2}, lends some support that it might be possible
to construct conserved mixed superenergy currents in this case as well.

\section*{Acknowledgments}

The author wishes to thank Göran Bergqvist and José Senovilla for
valuable comments and discussions. This work was partly carried out 
while visiting the Department of Theoretical Physics and History of Science at 
the University of the Basque Country.

\end{document}